\documentclass[aps,11pt]{revtex4}
\usepackage{amsmath}
\usepackage{bm}

\begin{document}

\preprint{}
\title{Distinguishing between $SU(5)$ and flipped $SU(5)$}
\vspace{2.0cm}
\author{Ilja Dorsner$^{1}$}
%\email{idorsner@ictp.trieste.it}
\author{Pavel Fileviez P\'erez$^{1,2}$}
%\email{fileviez@higgs.fis.puc.cl} 
\vspace{2.0cm}
\affiliation{$^{1}$The Abdus Salam International Centre
for Theoretical Physics\\
Strada Costiera 11, 34014 Trieste, Italy. \\ 
$^{2}$Pontificia Universidad Cat\'olica de Chile \\
Facultad de F{\'\i}sica, Casilla 306 \\
Santiago 22, Chile.}
\vspace{5.0cm}
%%%%%%%%%%%%%%%%%%%%%%%%%%%%%%%%%%%%%%%%%%%%%%%%%%%%%%%%%%%%%%%%%%%%
\begin{abstract}
We study in detail the $d=6$ operators for proton decay in the two
possible matter unification scenarios based on $SU(5)$ gauge symmetry. We
investigate the way to distinguish between these two 
scenarios. The dependence of the 
branching ratios for the two body decays on the fermion mixing is
presented in both cases. We point out the possibility to make 
a clear test of flipped $SU(5)$ through the decay 
channel $p \to \pi^+ \bar{\nu}$, and the ratio  
$\tau(p \to K^0 e^+_{\alpha}) / \tau(p \to \pi^0 e^+_{\alpha})$.
\end{abstract}
%%%%%%%%%%%%%%%%%%%%%%%%%%%%%%%%%%%%%%%%%%%%%%%%%%%%%%%%%%%%%%%%%%%%
\pacs{}
\maketitle
%%%%%%%%%%%%%%%%%%%%%%%%%%%%%%%%%%%%%%%%%%%%%%%%%%%%%%%%%%%%%%%%%%%%
\section{Introduction}
%%%%%%%%%%%%%%%%%%%%%%%%%%%%%%%%%%%%%%%%%%%%%%%%%%%%%%%%%%%%%%%%%%%%
Proton decay \cite{PatiSalam} is the most 
dramatic prediction of grand unified
theories, where quarks and leptons are at least
partially unified. Its signatures have been
extensively studied in various
theories~\cite{Weinberg,Zee,Sakai,Dimopoulos,Buras,Nathp1,Nathp2,Hisano1,Raby,Murayama,Goto} for many years. Recently, 
in the context of minimal supersymmetric $SU(5)$,
the predictions coming from both $d=5$ and $d=6$ operators have
been studied in order to understand if this model is ruled out \cite{Murayama,Goto}.  
Several solutions have been forwarded~\cite{Bajc1,Bajc2,Costa} 
to address this issue. This has renewed the interests of 
many groups to the important question of the proton
stability (for a review see \cite{Patireview}). 
Similar study, in the context of flipped $SU(5)$, has also been 
made \cite{Ellis3} concluding 
that the flipped model is out of trouble.
\\
There are several contributions to the decay of the proton. 
The $d=4$ and $d=5$ are the most important in supersymmetric scenarios. 
In a theory where matter-parity is conserved the $d=4$ are forbidden, while 
the $d=5$ operators can always be suppressed by choosing a 
particular Higgs sector \cite{Nath,Gomez,Babu}. The less model dependent
contributions are the $d=6$, which we study here in detail.  
\\
An extensive study of $d=6$ operators in the most general way 
in the context of $SU(5)$ and $SO(10)$ has been preformed in 
reference~\cite{Pavel}. There, it has
been pointed out that it is possible to make a clear test of any
grand unified theory with symmetric Yukawa couplings through the
decay channels into antineutrinos. However, the particular case of
flipped $SU(5)$ has not been studied taking into account the general
dependence on fermion mixing (for early analyses see
\cite{Barr,Ellis1,Ellis2}). With this work we seek to remedy that. Namely we 
investigate all $d=6$ proton decay operators in two different GUT models based
on $SU(5)$. We then confront the signatures of the two unifying
schemes pointing out the way to distinguish between them. We also
point out the way to test flipped $SU(5)$.
\\
The manuscript is organized as follows. In Section \ref{11} we
briefly review the key properties of both $SU(5)$ and flipped
$SU(5)$ unified theory. Section \ref{22} is devoted to the general
discussion of $d=6$ operators in both scenarios. In Section
\ref{33} we specify all the branching ratios for the independent
channels for proton decay. That section contains the main results
of our work. Finally, we conclude in the last section. The
Appendix contains useful decay rate formulas used throughout the
manuscript.
%%%%%%%%%%%%%%%%%%%%%%%%%%%%%%%%%%%%%%%%%%%%%%%%%%%%%%%%%%%%%%%%%%%%
\section{Matter unification based on $SU(5)$}
%%%%%%%%%%%%%%%%%%%%%%%%%%%%%%%%%%%%%%%%%%%%%%%%%%%%%%%%%%%%%%%%%%%%
\label{11}
The smallest special unitary group that contains the
Standard Model (SM) gauge group is $SU(5)$. The $SU(5)$ grand unified
theory \cite{Georgi,SUSYSU(5)} is an
anomaly free theory, where we have partial matter unification for each family in three
representation: ${\bf 10}$, $\overline{{\bf 5}}$ and ${\bf 1}$. The
singlet is identify with the right handed neutrino. 
In the SM language we have: ${\bf 10} = ({\bf 3},{\bf 2},1/3) \oplus
(\overline{{\bf 3}},{\bf 1},-4/3) \oplus ({\bf 1},{\bf
1},2)=(Q,u^C,e^C)$, $\overline{{\bf 5}} = (\overline{{\bf 3}},{\bf
1},2/3) \oplus ({\bf 1},{\bf 2},-1) =(d^C,L)$, and ${\bf 1} =
({\bf 1},{\bf 1},0)=\nu^C$, where $Q=(u,d)$ and $L= (\nu,e)$.
The off-diagonal part of the gauge fields residing in the ${\bf 24}$ of $SU(5)$ is 
composed of bosons $(X,Y)=({{\bf 3}},{\bf 2},5/3)$ and 
their conjugates, which mediate proton decay. $X$ and
$Y$ fields have electric charge $4/3$ and $1/3$, respectively.
\\
The electric charge is a generator of conventional $SU(5)$. However, 
it is possible to embed the electric charge in such a manner 
that it is a linear combination of the generators 
operating in both $SU(5)$ and an
extra $U(1)$, and still reproduce the SM charge assignment. 
This is exactly what is done in a flipped 
$SU(5)$~\cite{DeRujula,Barr,Derendinger,Antoniadis}. 
The matter now unifies in a different manner, which can be obtained
from the $SU(5)$ assignment by a flip: $d^C \leftrightarrow u^C$,
$e^C \leftrightarrow \nu^C$, $u \leftrightarrow d$ and
$\nu \leftrightarrow e$. 
In the case of flipped $SU(5)$ the gauge bosons responsible for
proton decay are: $(X', Y')=({{\bf 3}},{\bf 2},-1/3)$. 
The electric charge of $Y'$ is $-2/3$, while
$X'$ has the same charge as $Y$. Since the gauge sector and the
matter unification differ from $SU(5)$ case, the proton decay
predictions are also different~\cite{Barr}.
\\
Flipped $SU(5)$ is well motivated from string theory scenarios, since
we do not need large representations to achieve the GUT symmetry
breaking \cite{Antoniadis}.
Another nice feature of flipped $SU(5)$ is that the dangerous $d=5$
operators are suppressed due to an extremely economical missing
partner mechanism. This allows us to concentrate our attention to the
gauge $d=6$ contributions. 
\\
We next analyze the possibility to test two realistic grand
unified theories: the $SU(5)$ and flipped $SU(5)$ theory. 
We make an analysis of the operators in each theory,
and study the physical parameters entering in the predictions for
proton decay. We do not commit to any particular model for fermion
masses, in order to be sure that we can test the grand unification
idea. 
%
%%%%%%%%%%%%%%%%%%%%%%%%%%%%%%%%%%%%%%%%%%%%%%%%%%%%%%%%%%%%%%%%%%%%
\section{d=6 operators}
%%%%%%%%%%%%%%%%%%%%%%%%%%%%%%%%%%%%%%%%%%%%%%%%%%%%%%%%%%%%%%%%%%%%
\label{22}
In the Georgi-Glashow $SU(5)$ matter unification case, the gauge $d=6$
operators contributing to the decay of the proton are \cite{Weinberg,Zee,Sakai}:
\begin{subequations}
\label{SU5}
\begin{eqnarray}
\label{O1} \textit{O}^{B-L}_{SU(5)}&=& k^2_1 \ \epsilon_{ijk} \
\epsilon_{\alpha \beta} \ \overline{u_{i a}^C} \ \gamma^{\mu} \
Q_{j \alpha a}   \
\overline{e_b^C} \ \gamma_{\mu} \ Q_{k \beta b},\\
\label{O2} \textit{O}^{B-L}_{SU(5)}&=& k^2_1 \ \epsilon_{ijk} \
\epsilon_{\alpha \beta} \ \overline{u_{i a}^C} \ \gamma^{\mu} \
Q_{j \alpha a}   \ \overline{d^C_{k b}} \ \gamma_{\mu} \ L_{\beta
b}.
\end{eqnarray}
\end{subequations}
On the other hand, flipped $SU(5)$ matter unification yields:
\begin{subequations}
\label{flippedSU5}
\begin{eqnarray}
\label{O3} \textit{O}^{B-L}_{SU(5)'}&=& k^2_2 \ \epsilon_{ijk} \
\epsilon_{\alpha \beta} \ \overline{d_{i a}^C} \ \gamma^{\mu} \
Q_{j \beta a}   \
\overline{u_{k b}^C} \ \gamma_{\mu} \ L_{\alpha b}, \\
\label{O4} \textit{O}^{B-L}_{SU(5)'}&=& k^2_2 \ \epsilon_{ijk} \
\epsilon_{\alpha \beta} \ \overline{d^C_{i a}} \ \gamma^{\mu} \
Q_{j \beta a}   \ \overline{\nu_b^C} \ \gamma_{\mu} \ Q_{k \alpha
b}.
\end{eqnarray}
\end{subequations}
In the above expressions $k_1= g_5 M^{-1}_{(X,Y)}$, and $k_2= g'_5
{M^{-1}_{(X',Y')}}$, where $M_{(X,Y)}$ ($M_{(X',Y')}$) $\sim
M_{GUT}\approx 10^{16}$\,GeV and $g_5$ ($g'_5$) are the masses of
the superheavy gauge bosons and the couplings at the GUT scale in
$SU(5)$ (flipped $SU(5)$) case. $i$, $j$ and $k$ are the color
indices, $a$ and $b$ are the family indices, and $\alpha, \beta
=1,2$.
\\
In these theories the diagonalization of the Yukawa matrices
is given by the following bi-unitary transformations:
\begin{eqnarray}
\label{YUd}
U^T_C \ Y_U \ U &=& Y_U^{\textrm{diag}},\\
\label{YDd}
D^T_C \ Y_D \ D &=& Y_D^{\textrm{diag}},\\
\label{YEd} E^T_C \ Y_E \ E &=& Y_E^{\textrm{diag}}.
\end{eqnarray}
Using the operators listed in Eqs.~(\ref{SU5}), the effective
operators for each decay channel in the $SU(5)$ case upon Fierz
transformation take the following form in the physical basis \cite{Pavel}:
\begin{subequations}
\begin{eqnarray}
\label{1Oec1} \textit{O}(e_{\alpha}^C, d_{\beta})_{SU(5)}&=&
c(e^C_{\alpha}, d_{\beta})_{SU(5)} \ \epsilon_{ijk} \
\overline{u^C_i} \ \gamma^{\mu} \ u_j \ \overline{e^C_{\alpha}} \
\gamma_{\mu} \ d_{k \beta}, \\
\label{1Oe} \textit{O}(e_{\alpha}, d^C_{\beta})_{SU(5)}&=&
c(e_{\alpha}, d^C_{\beta})_{SU(5)} \ \epsilon_{ijk} \
\overline{u^C_i} \ \gamma^{\mu} \ u_j \ \overline{d^C_{k \beta}} \
\gamma_{\mu} \ e_{\alpha},\\
\label{1On} \textit{O}(\nu_l, d_{\alpha}, d^C_{\beta} )_{SU(5)}&=&
c(\nu_l, d_{\alpha}, d^C_{\beta})_{SU(5)} \ \epsilon_{ijk} \
\overline{u^C_i} \ \gamma^{\mu} \ d_{j \alpha} \ \overline{d^C_{k
\beta}} \ \gamma_{\mu} \ \nu_l,\\
\label{1OnC} \textit{O}(\nu_l^C, d_{\alpha}, d^C_{\beta}
)_{SU(5)}&=& c(\nu_l^C, d_{\alpha}, d^C_{\beta})_{SU(5)} \
\epsilon_{ijk} \ \overline{d_{i \beta}^C} \ \gamma^{\mu} \ u_j \
\overline{\nu_l^C} \ \gamma_{\mu} \ d_{k \alpha},
\end{eqnarray}
\end{subequations}
where
\begin{subequations}
\label{coeffSU5}
\begin{eqnarray}
\label{1cec} c(e^C_{\alpha}, d_{\beta})_{SU(5)}&=& k_1^2 \left[
V^{11}_1 V^{\alpha \beta}_2 + ( V_1 V_{UD})^{1
\beta}( V_2 V^{\dagger}_{UD})^{\alpha 1}\right], \\
\label{1ce} c(e_{\alpha}, d_{\beta}^C)_{SU(5)} &=& k_1^2 V^{11}_1
V^{\beta \alpha}_3, \\
\label{1cnu} c(\nu_l, d_{\alpha}, d^C_{\beta})_{SU(5)}&=& k_1^2 (
V_1 V_{UD} )^{1 \alpha} ( V_3 V_{EN})^{\beta l}, \ \textrm{$\alpha=1$ or $\beta=1$}\\
\label{1cnuc} c(\nu_l^C, d_{\alpha}, d^C_{\beta})_{SU(5)}&=&0.
\end{eqnarray}
\end{subequations}
In the case of flipped $SU(5)$ (see Eqs.~(\ref{flippedSU5})) the
effective operators are
\begin{subequations}
\begin{eqnarray}
\label{2Oec1} \textit{O}(e_{\alpha}^C, d_{\beta})_{SU(5)'}&=&
c(e^C_{\alpha}, d_{\beta})_{SU(5)'} \ \epsilon_{ijk} \
\overline{u^C_i} \ \gamma^{\mu} \ u_j \ \overline{e^C_{\alpha}} \
\gamma_{\mu} \ d_{k \beta}, \\
\label{2Oe} \textit{O}(e_{\alpha}, d^C_{\beta})_{SU(5)'}&=&
c(e_{\alpha}, d^C_{\beta})_{SU(5)'} \ \epsilon_{ijk} \
\overline{u^C_i} \ \gamma^{\mu} \ u_j \ \overline{d^C_{k \beta}} \
\gamma_{\mu} \ e_{\alpha},\\
\label{2On} \textit{O}(\nu_l, d_{\alpha}, d^C_{\beta}
)_{SU(5)'}&=& c(\nu_l, d_{\alpha}, d^C_{\beta})_{SU(5)'} \
\epsilon_{ijk} \ \overline{u^C_i} \ \gamma^{\mu} \ d_{j \alpha}
\ \overline{d^C_{k \beta}} \ \gamma_{\mu} \ \nu_l, \\
\label{2OnC} \textit{O}(\nu_l^C, d_{\alpha}, d^C_{\beta}
)_{SU(5)'}&=& c(\nu_l^C, d_{\alpha}, d^C_{\beta})_{SU(5)'} \
\epsilon_{ijk} \ \overline{d_{i \beta}^C} \ \gamma^{\mu} \ u_j \
\overline{\nu_l^C} \ \gamma_{\mu} \ d_{k \alpha},
\end{eqnarray}
\end{subequations}
where
\begin{subequations}
\label{coeffflippedSU5}
\begin{eqnarray}
\label{2cec} c(e^C_{\alpha}, d_{\beta})_{SU(5)'}&=&0,\\
 \label{2ce}
c(e_{\alpha}, d_{\beta}^C)_{SU(5)'} &=&
k_2^2 (V_4 V^{\dagger}_{UD} )^{\beta 1} ( V_1 V_{UD} V_4^{\dagger} V_3)^{1 \alpha},\\
\label{2cnu} c(\nu_l, d_{\alpha}, d^C_{\beta})_{SU(5)'}&=&  k_2^2
V_4^{\beta \alpha}( V_1 V_{UD}
V^{\dagger}_4 V_3 V_{EN})^{1l},  \ \textrm{$\alpha=1$ or $\beta=1$}\\
\label{2cnuc} c(\nu_l^C, d_{\alpha}, d^C_{\beta})_{SU(5)'}&=&
k_2^2 \left[ ( V_4 V^{\dagger}_{UD} )^{\beta
 1} ( U^{\dagger}_{EN} V_2)^{l \alpha }+ V^{\beta \alpha}_4
 (U^{\dagger}_{EN} V_2 V^{\dagger}_{UD})^{l1}\right], \ \textrm{$\alpha=1$ or $\beta=1$}.
\end{eqnarray}
\end{subequations}
We use the subscripts $SU(5)$ and $SU(5)'$ to distinguish the two scenarios.
The mixing matrices $V_1= U_C^{\dagger} U$, $V_2=E_C^{\dagger}D$,
$V_3=D_C^{\dagger}E$, $V_4=D_C^{\dagger} D$,
$V_{UD}=U^{\dagger}D$, $V_{EN}=E^{\dagger}N$ and $U_{EN}=
E_C^{\dagger} N_C$. The quark mixing is given by
$V_{UD}=U^{\dagger}D=K_1 V_{CKM} K_2$, where $K_1$ and $K_2$ are
diagonal matrices containing three and two phases, respectively.
The leptonic mixing $V_{EN}=K_3 V^D_l K_4$ in case of Dirac
neutrino, or $V_{EN}=K_3 V^M_l$ in the Majorana case. $V^D_l$ and
$V^M_l$ are the leptonic mixing matrices at low energy in the Dirac and
Majorana case, respectively.
\\
Notice that in general to predict the lifetime of the proton in
$SU(5)$, due to the presence of $d=6$ operators, we have to know $k_1$,
$V^{1b}_1$, $V_2$, $V_3$, while in flipped $SU(5)$ we have to know
$k_2$, $V^{1b}_1$, $V_3$, $V_4$ and $U_{EN}$. In addition we have to
know three diagonal matrices containing CP violating phases, $K_1$,
$K_2$ and $K_3$, in the case that the neutrino is Majorana. In the
Dirac case there is an extra matrix with two more phases.
\\
From the above equations, we see that there are no decays into 
$\nu^C$ in $SU(5)$, and in flipped $SU(5)$ into $e^C$, since 
these are singlets in the corresponding scenarios.
%
%%%%%%%%%%%%%%%%%%%%%%%%%%%%%%%%%%%%%%%%%%%%%%%%%%%%%%%%%%%%%%%%%%%%%%%%%%%
\section{Flipped $SU(5)$ versus $SU(5)$}
%%%%%%%%%%%%%%%%%%%%%%%%%%%%%%%%%%%%%%%%%%%%%%%%%%%%%%%%%%%%%%%%%%%%%%%%%%%
\label{33} There are only seven independent relations for all
coefficients of the gauge d=6 operators contributing to nucleon
decay \cite{Pavel}. Therefore, if we want to test a grand unified
theory, the number of physical quantities entering in the proton
decay amplitude must be less than that. This is important to know
in order to see if it is possible to test a GUT scenario.
\\
Since we cannot distinguish between the neutrino flavors in the
proton decay experiments, in order to compute the branching ratios
into antineutrinos we have to sum over all of them. Using the
expressions in the Appendix, and the following relations:
\\
\begin{subequations}
\begin{eqnarray}
\label{sumSU(5)} \sum_{l=1}^3 c(\nu_l, d_{\alpha},
d^C_{\beta})_{SU(5)}^* c(\nu_l, d_{\gamma},
d^C_{\delta})_{SU(5)}&=& k_1^4 (V_1^* V_{UD}^*)^{1 \alpha} (V_1
V_{UD})^{1 \gamma} \delta^{\beta \delta},
\\
\nonumber\\
\label{sumflippedSU(5)} \sum_{l=1}^3 c(\nu_l, d_{\alpha},
d^C_{\beta})_{SU(5)'}^* c(\nu_l, d_{\gamma},
d^C_{\delta})_{SU(5)'}&=& k_2^4 (V_4^*)^{\beta \alpha} V_4^{\delta \gamma},
\end{eqnarray}
\end{subequations}
we can write down the ratios between the lifetimes in both
theories for the decays into antineutrinos. They are:
\\
\begin{subequations}
\label{R1}
\begin{eqnarray}
\label{ratio1} \frac{\tau(p \to K^+\bar{\nu})^{SU(5)'}}{\tau(p \to
K^+\bar{\nu})^{SU(5)}}&=&\frac{k_1^4}{k_2^4} \frac{A_1^2
\left|(V_1 K_1 V_{CKM})^{11}\right|^2+A_2^2 \left|(V_1 K_1
V_{CKM})^{12}\right|^2}{A_1^2 \left|V_4^{21}\right|^2 +A_2^2
\left|V_4^{12}\right|^2+A_1 A_2 \left((V^*_4)^{21}
V_4^{12}+(V^*_4)^{12} V_4^{21}\right)},
\\
\nonumber \\
\label{ratio2} \frac{\tau(p \to \pi^+\bar{\nu})^{SU(5)'}}{\tau(p
\to \pi^+\bar{\nu})^{SU(5)}}&=&\frac{k_1^4}{k_2^4}
\frac{\left|(V_1 K_1
V_{CKM})^{11}\right|^2}{\left|V_4^{11}\right|^2},\\
\nonumber \\
\label{ratio5} \frac{\tau(n \to K^0\bar{\nu})^{SU(5)'}}{\tau(n \to
K^0\bar{\nu})^{SU(5)}}&=&\frac{k_1^4}{k_2^4} \frac{A_3^2
\left|(V_1 K_1 V_{CKM})^{11}\right|^2+A_2^2 \left|(V_1 K_1
V_{CKM})^{12}\right|^2}{A_3^2 \left|V_4^{21}\right|^2 +A_2^2
\left|V_4^{12}\right|^2+A_3 A_2 \left((V^*_4)^{21}
V_4^{12}+(V^*_4)^{12} V_4^{21}\right)},
\end{eqnarray}
\end{subequations}
where
\begin{subequations}
\begin{eqnarray}
A_1&=& \frac{2 m_p }{3 m_B} D,\\
A_2&=& 1 + \frac{m_p}{3 m_B}(D + 3F),\\
A_3&=& 1 + \frac{m_n}{3 m_B}(D-3F).
\end{eqnarray}
\end{subequations}
The same procedure can be done for the decays into charged leptons:
\begin{eqnarray}
\label{ratio3} \frac{\tau(p \rightarrow \pi^0
e_{\beta}^+)^{SU(5)'}}{\tau(p \rightarrow \pi^0
e_{\beta}^+)^{SU(5)}}&=&\frac{k_1^4}{k_2^4} \frac{\left|V^{11}_1
V_3^{1\beta} \right|^2+\left|V^{11}_1 V_2^{\beta 1}+(V_1 K_1
V_{CKM} K_2)^{11} (V_2 K^*_2 V^\dagger_{CKM} K^*_1)^{\beta 1}
\right|^2}{\left|(V_4 K^*_2 V^\dagger_{CKM} K^*_1)^{11}(V_1 K_1
V_{CKM} K_2 V^\dagger_4 V_3)^{1\beta}\right|^2}
\\ \nonumber\\
\label{ratio4} \frac{\tau(p \rightarrow K^0
e_{\beta}^+)^{SU(5)'}}{\tau(p \rightarrow K^0
e_{\beta}^+)^{SU(5)}}&=&\frac{k_1^4}{k_2^4} \frac{\left|V^{11}_1
V_3^{2\beta} \right|^2+\left|V^{11}_1 V_2^{\beta 2}+(V_1 K_1
V_{CKM} K_2)^{12} (V_2 K^*_2 V^\dagger_{CKM} K^*_1)^{\beta 1}
\right|^2}{\left|(V_4 K^*_2 V^\dagger_{CKM} K^*_1)^{21}(V_1 K_1
V_{CKM} K_2 V^\dagger_4 V_3)^{1\beta}\right|^2} 
\end{eqnarray}
\\
Eqs.~(\ref{R1}), (\ref{ratio3}), and (\ref{ratio4}) are the 
most general equations that we could 
write in the two scenarios and will help in future to
distinguish between them if proton decay is found. In other words,
for a given model of fermion masses, using the above equations we
could see the difference in the predictions for proton decay.
Unfortunately, as one can appreciate, the branching ratios depend
on too many unknown factors, including the new CP 
violating phases. (These, in principle, could be defined in a
particular model for CP violation.) Therefore it is impossible to
test those scenarios in general through the decay of the proton unless
we known the flavor structure of the SM fermions.
\\
Since we cannot make clear predictions in the most general case, let
us consider special cases in these two matter unification
scenarios based on $SU(5)$ and compare them.
%
%%%%%%%%%%%%%%%%%%%%%%%%%%%%%%%%%%%%%%%%%%%%%%%%%%%%%%%%%%%%%%%%%%
\subsection{$SU(5)$ with $Y_U = Y_U^T$}
%%%%%%%%%%%%%%%%%%%%%%%%%%%%%%%%%%%%%%%%%%%%%%%%%%%%%%%%%%%%%%%%%%
%
In $SU(5)$, if $Y_U=Y_U^T$, we have $U_C = U K_u$, where $K_u$ is a
diagonal matrix containing three CP violating phases. Therefore we get:
\begin{eqnarray}
\label{sumSU(5)1} \sum_{l=1}^3 c(\nu_l, d_{\alpha},
d^C_{\beta})_{SU(5)}^* c(\nu_l, d_{\gamma},
d^C_{\delta})_{SU(5)}&=& k_1^4 (V_{CKM}^*)^{1 \alpha}
(K_2^*)^{\alpha \alpha} (V_{CKM})^{1 \gamma} K_2^{\gamma \gamma}
\delta^{\beta \delta}.
\end{eqnarray}
In this case, as has been shown \cite{Pavel}, the clean
channels, i.e., the channels that we have to look to test this
scenario, are:
\begin{subequations}
\begin{eqnarray}
\label{xxx1} \Gamma(p \to K^+\bar{\nu})&=& k_1^4 \left[ A^2_1
|V_{CKM}^{11}|^2+A^2_2 |V_{CKM}^{12}|^2\right] C_1,
\\
\label{xxx2} \Gamma(p \rightarrow \pi^+ \bar{\nu}) &=& k_1^4
\left|V_{CKM}^{11}\right|^2 C_2,
\end{eqnarray}
\end{subequations}
where \begin{subequations}
\begin{eqnarray}\label{a}
C_1&=&\frac{(m_p^2-m_K^2)^2}{8\pi m_p^3 f_{\pi}^2} A_L^2
\left|\alpha\right|^2,
\\
\label{b} C_2&=&\frac{m_p}{8\pi f_{\pi}^2}  A_L^2 \left|\alpha
        \right|^2 (1+D+F)^2.
\end{eqnarray}
\end{subequations}
Notice that we have two expressions for $k_1$, which are independent
of the unknown mixing matrices and the CP violating phases.
Therefore it is possible to test $SU(5)$ grand unified theory 
with symmetric up Yukawa matrices through these two channels~\cite{Pavel}. Notice
that these results are valid for any unified model based on $SU(5)$ 
with $Y_U= Y_U^T$. For example, this includes the case of minimal SUSY $SU(5)$ with 
two extra Higgses in the fundamental and antifundamental
representations. The case of modified missing doublet SUSY $SU(5)$
model \cite{Masiero,Mohapatra} is also included in our analysis. 
%
%%%%%%%%%%%%%%%%%%%%%%%%%%%%%%%%%%%%%%%%%%%%%%%%%%%%%%%%%%%%%%%%%%
\subsection{Renormalizable flipped $SU(5)$}
%%%%%%%%%%%%%%%%%%%%%%%%%%%%%%%%%%%%%%%%%%%%%%%%%%%%%%%%%%%%%%%%%%
%
In renormalizable flipped $SU(5)$ we have $Y_D=Y_D^T$, so 
$D_C = D K_d$, where $K_d$ is a diagonal matrix containing three CP
violating phases. In this case the coefficients entering in the proton
decay predictions are:
\\
\begin{subequations}
\begin{eqnarray}
\label{sumflippedSU(5)1} \sum_{l=1}^3 c(\nu_l, d_{\alpha},
d^C_{\beta})_{SU(5)'}^*  \ c(\nu_l, d_{\gamma},
d^C_{\delta})_{SU(5)'} & = & k_2^4 K_d^{\beta \beta} \delta^{\beta
\alpha} (K_d^*)^{\delta \delta} \delta^{\delta \gamma},
\\ \nonumber\\
\left|c(e_{\alpha}, d^C_{\beta})\right|^2=k_2^4 \left|V_{CKM}^{1
\beta}\right|^2 \left|(V_1 V_{UD} V^\dagger_4 V_3)^{1
\alpha}\right|^2 & = & k_2^4 \left|V_{CKM}^{1 \beta}\right|^2
\left|(U_C^\dagger E)^{1 \alpha}\right|^2,
\end{eqnarray}
\end{subequations}
\\
Using these equations we get the following relations:
\\
\begin{subequations}
\begin{eqnarray}
\label{xxx5}
\Gamma(p \rightarrow \pi^+ \bar{\nu})&=& k_2^4 \ C_2,
\\ \nonumber\\
\label{flippedpi0}
\Gamma(p \rightarrow \pi^0 e_{\alpha}^+)&=&\frac{1}{2} \ \Gamma(p
\rightarrow \pi^+\bar{\nu}) \ \left|V_{CKM}^{1 1}\right|^2
\left|(U_C^\dagger E)^{1 \alpha}\right|^2,
\\ \nonumber\\
\label{flippedK0}
\frac{\Gamma (p \to K^0 e_{\alpha}^+)}
{\Gamma (p \to \pi^0 e_{\alpha}^+)}
        &=& 2 \frac{C_3}{C_2} \ \frac{\left|V_{CKM}^{12}\right|^2}
{\left|V_{CKM}^{11}\right|^2},
\end{eqnarray}
\end{subequations}
\\
where:
\begin{equation}
C_3 = \frac{(m_p^2-m_K^2)^2}{8 \pi f_\pi^2 m_p^3}  A_L^2
        \left|\alpha\right|^2 \left[1+{\frac{m_p}{m_B}} (D-F)\right]^2.
\end{equation}
\\
Notice that in this case, $\Gamma(p \to K^+\bar{\nu})= 0$, and $\Gamma(n
\to K^0 \bar{\nu}) =0$. In Eq.~(\ref{flippedK0}) we assume
$(U_C^\dagger E)^{1 \alpha} \neq 0$.
\\
We can say that the renormalizable flipped $SU(5)$ can
be verified by looking at the channel $p \to \pi^+ \bar{\nu}$, and
using the correlation stemming from Eq.~(\ref{flippedK0}). 
This is a nontrivial result and can help us to test 
this scenario, if proton decay 
is found in the next generation of experiments. It is one of 
the main results of this work. If this channel is measured, we can 
know the predictions for decays into charged leptons using 
Eq.~(\ref{flippedpi0}) for a given model for 
fermion masses. Therefore it is possible to differentiate between
different fermion mass models.
\\
Note the difference between Eqs.~(\ref{xxx2}) and (\ref{xxx5}); there 
appears a suppression factor for the channel $p \to \pi^+ {\overline{\nu}}$ 
in the case of $SU(5)$.   
\\
Since the nucleon decays into $K$ mesons are absent in the case of flipped
$SU(5)$, that is an independent way to distinguish this model from
$SU(5)$, where these channels are always present.
%
%%%%%%%%%%%%%%%%%%%%%%%%%%%%%%%%%%%%%%%%%%%%%%%%%%%%%%%%%%%%%%%%%
\section{Conclusions}
%%%%%%%%%%%%%%%%%%%%%%%%%%%%%%%%%%%%%%%%%%%%%%%%%%%%%%%%%%%%%%%%%
%
We have investigated in model independent way 
the predictions coming from the gauge $d=6$ 
operators in the two possible matter unification scenarios based 
on $SU(5)$ gauge symmetry. 
We write down the most general ratios between the lifetimes in 
$SU(5)$ and flipped $SU(5)$ theory for each 
channel, providing the way to distinguish between them. 
We find that in general it is very difficult to
test flipped $SU(5)$. However, in the case of renormalizable
flipped $SU(5)$ model, the decay channel $ p \to \pi^+ {\overline
\nu}$, which is a clean channel, and the ratio 
$\tau(p \to K^0 e^+_{\alpha}) / \tau(p \to \pi^0 e^+_{\alpha})$ 
could be used to test this theory. 
If the decay of the proton is found in future, our results
will be useful to analyze the predictions in these theories. 
\begin{acknowledgments}
P.F.P thanks Goran Senjanovi\'c for comments, and the High Energy
Section of the ICTP for their hospitality and support. This work was
supported in part by CONICYT/FONDECYT under contract $N^{\underline 0}
\ 3050068$. 
\end{acknowledgments}
\appendix*
%%%%%%%%%%%%%%%%%%%%%%%%%%%%%%%%%%%%%%%%%%%%%%%%%%%%%%%%%%
\section{}
%%%%%%%%%%%%%%%%%%%%%%%%%%%%%%%%%%%%%%%%%%%%%%%%%%%%%%%%%%
Using the chiral Lagrangian techniques (see reference
\cite{Chadha}), the decay rate of the different channels due to
the presence of the gauge $d=6$ operators are given by:
\begin{eqnarray}
\label{A1} \Gamma(p \to K^+\bar{\nu})
        &=& \frac{(m_p^2-m_K^2)^2}{8\pi m_p^3 f_{\pi}^2} A_L^2
\left|\alpha\right|^2 \nonumber\\
&&\times \sum_{i=1}^3 \left|\frac{2m_p}{3m_B}D \ c(\nu_i, d, s^C)
+ [1+\frac{m_p}{3m_B}(D+3F)] c(\nu_i,s, d^C)\right|^2,\\
\label{A2} \Gamma(p \to \pi^+\bar{\nu})
        &=&\frac{m_p}{8\pi f_{\pi}^2}  A_L^2 \left|\alpha
        \right|^2 (1+D+F)^2
\sum_{i=1}^3 \left| c(\nu_i, d, d^C) \right|^2,\\
\label{A3} \Gamma(p \to \eta e_{\beta}^+)
        &=& {\frac{(m_p^2-m_\eta^2)^2}{ 48 \pi f_\pi^2 m_p^3}}
A_L^2 \left|\alpha \right|^2 (1+D-3 F)^2 \{ \left|
c(e_{\beta},d^C)\right|^2 +
\left|c(e^C_{\beta}, d)\right|^2 \},\\
\label{A4} \Gamma (p \to K^0 e_{\beta}^+)
        &=& {\frac{(m_p^2-m_K^2)^2}{8 \pi f_\pi^2 m_p^3}}  A_L^2
        \left|\alpha\right|^2 [1+{\frac{m_p}{m_B}} (D-F)]^2
\{ \left|c(e_{\beta},s^C)\right|^2 + \left|c(e^C_{\beta},s)\right|^2\},\\
\label{A5} \Gamma(p \rightarrow \pi^0 e_{\beta}^+)
           &=& \frac{m_p}{16\pi f_{\pi}^2} A_L^2
           \left|\alpha\right|^2
        (1+D+F)^2 \{ \left|c(e_{\beta},d^C)\right|^2 +
        \left|c(e^C_{\beta},d)\right|^2 \},
\\
\label{A6} \Gamma(n \rightarrow K^0 \overline\nu)&=&
\frac{(m_n^2-m_K^2)^2}{8
\pi m_n^3 f_\pi^2} A_L^2 \left|\alpha\right|^2 \nonumber\\
&& \times \sum_{i=1}^3 \left|c(\nu_i,d,s^C) [1+\frac{m_n}{3 m_B}
(D-3 F)]-c(\nu_i,s,d^C)[1+\frac{m_n}{3 m_B}(D+3 F)]\right|^2\\
\label{A7} \Gamma(n \rightarrow \pi^0
\overline\nu)&=&\frac{m_n}{16 \pi f_\pi^2} A_L^2
        \left|\alpha\right|^2 (1+D+F)^2 \sum_{i=1}^3 \left|c(\nu_i, d,
        d^C)\right|^2,\\
\label{A8} \Gamma\rightarrow(n \to \eta \overline\nu)&=&
\frac{(m_n^2-m_\eta^2)^2}{ 48 \pi m_n^3 f_\pi^2} A_L^2
\left|\alpha\right|^2 (1+D-3 F)^2
\sum_{i=1}^3 \left|c(\nu_i, d, d^C)\right|^2, \\
\label{A9} \Gamma(n \rightarrow \pi^- e^+_{\beta})&=&\frac{ m_n}{
8 \pi f_\pi^2} A_L^2
        \left|\alpha\right|^2 (1+D+F)^2
        \{\left|c(e_{\beta}, d^C)\right|^2 +
        \left|c(e^C_{\beta},d)\right|^2\}.
\end{eqnarray}
In the above equations $m_B$ is an average Baryon mass satisfying
$m_B \approx m_\Sigma \approx m_\Lambda$, $D$, $F$ and $\alpha$
are the parameters of the chiral lagrangian, and all other
notation follows \cite{Chadha}. Here all coefficients of
four-fermion operators are evaluated at $M_Z$ scale. $A_L$ takes
into account renormalization from $M_Z$ to 1 GeV. $\nu_i= \nu_e,
\nu_{\mu}, \nu_{\tau}$ and $e_{\beta}= e, \mu$.
%%%%%%%%%%%%%%%%%%%%%%%%%%%%%%%%%%%%%%%%%%%%%%%%%%%%%%%%%%%%%%%%%
%%%%%%%%%%%%%%%%%%%%%%%%%%%%%%%%%%%%%%%%%%%%%%%%%%%%%%%%%%%%%%%%%

\end{document}